\documentclass{elsart}

\usepackage{epsfig}

\usepackage{amssymb}

\begin{document}

\begin{frontmatter}



% declarations for front matter
\title{Near Muon Range Detector for the K2K Experiment\\
	- Construction and Performance -}

%%%%% mbox each name so the initials don't break away along lines.
\author[KEK]{\mbox{T. Ishii}\corauthref{cor}},
\author[Kyoto]{\mbox{T. Inagaki}},
\author[UCI]{\mbox{J. Breault}},
\author[KEK]{\mbox{T. Chikamatsu}\thanksref{CHIKAnow}},
\author[CNU]{\mbox{J. H. Choi}},
\author[Tohoku]{\mbox{T. Hasegawa}},
\author[KEK]{\mbox{Y. Hayato}},
\author[KEK]{\mbox{T. Ishida}},
\author[CNU]{\mbox{H. I. Jang}\thanksref{JANGnow}},
\author[CNU]{\mbox{J. S. Jang}},
\author[CNU]{\mbox{E. M. Jeong}},
\author[Kyoto]{\mbox{I. Kato}},
\author[Hawaii]{\mbox{A. Kibayashi}},
\author[SNU]{\mbox{B. J. Kim}},
\author[SNU]{\mbox{H. I. Kim}},
\author[CNU]{\mbox{J. Y. Kim}},
\author[SNU]{\mbox{S. B. Kim}},
\author[KEK]{\mbox{T. Kobayashi}},
\author[UCI]{\mbox{W. Kropp}},
\author[CNU]{\mbox{H. K. Lee}},
\author[KEK]{\mbox{S. B. Lee}},
\author[CNU]{\mbox{I. T. Lim}},
\author[CNU]{\mbox{K. J. Ma}},
\author[Tohoku]{\mbox{T. Maruyama}\thanksref{KEKnow}},
\author[UCI]{\mbox{S. Mine}},
\author[KEK]{\mbox{K. Nakamura}},
\author[Niigata]{\mbox{M. Nakamura}},
\author[Okayama]{\mbox{I. Nakano}},
\author[Kyoto]{\mbox{K. Nishikawa}},
\author[KEK]{\mbox{Y. Oyama}},
\author[Dongshin]{\mbox{M. Y. Pac}},
\author[KEK]{\mbox{A. Sakai}},
\author[KEK]{\mbox{M. Sakuda}},
\author[KEK]{\mbox{K. Shiino}},
\author[KEK]{\mbox{K. Tauchi}},
\author[KEK]{\mbox{V. Tumakov}},
\author[SNU]{\mbox{J. Yoo}},
\author[CNU]{\mbox{S. Y. You}}

%%%%% Included file of addresses. 
%%%%% This gets spliced into the main file before submission.

\address[KEK]{Institute of Particle and Nuclear Studies, KEK, 
Tsukuba, Ibaraki 305-0801, JAPAN }

\address[Kyoto]{Department of Physics, Kyoto University, 
Kyoto 606-8502, JAPAN}

\address[UCI]{Department of Physics and Astronomy, University of 
California, Irvine, Irvine, CA 92697-4575, USA }

\address[CNU]{Department of Physics, Chonnam National University,
Kwangju 500-757, KOREA}

\address[Tohoku]{Research Center for Neutrino Science, 
Tohoku University, Sendai, Miyagi 980-8578, JAPAN}

\address[Hawaii]{Department of Physics and Astronomy, 
University of Hawaii, Honolulu, HI 96822, USA}

\address[SNU]{Department of Physics, Seoul National University,
Seoul 151-742, KOREA}

\address[Niigata]{Department of Physics, Niigata University,
Niigata, Niigata 950-2181, JAPAN}

\address[Okayama]{Department of Physics, Okayama University, 
Okayama, Okayama 700-8530,  JAPAN}

\address[Dongshin]{Department of Physics, Dongshin 
University, Naju 520-714, KOREA}

\corauth[cor]{Corresponding author. Tel.: +81 298 64 5425; 
fax: +81 298 64 7831; e-mail address: takanobu.ishii@kek.jp.}
% fax: +81 298 64 7831; \ead{takanobu.ishii@kek.jp (T. Ishii).}}

\thanks[CHIKAnow]{Present address: Miyagi Gakuin Women's College, Sendai, 
Miyagi 981-8557, JAPAN.}

\thanks[JANGnow]{Present address: Department of Civil Engineering, 
Seokang College, Kwangju, 500-742, KOREA.}

\thanks[KEKnow]{Present address: Institute of Particle and Nuclear Studies, 
KEK, Tsukuba, Ibaraki 305-0801, JAPAN }

\collab{K2K MRD Group}

\newpage

\begin{abstract}

A muon range detector~(MRD) has been constructed as a near detector for the 
KEK-to-Kamioka long-baseline neutrino experiment~(K2K).  
It monitors the neutrino beam properties at the near site 
by measuring the energy, angle and production point of muons 
produced by charged-current neutrino interaction. 
The detector has been working stably since the start of the K2K 
experiment. 
\end{abstract}

\begin{keyword}
muon \sep range \sep K2K \sep drift tube \sep neutrino \sep long baseline
\PACS 14.60.Pq \sep 14.60.Lm \sep 96.40.Tv
\end{keyword}
\end{frontmatter}

% main text

\section{Introduction}

The KEK-to-Kamioka long-baseline neutrino experiment~(K2K)\cite{K2K} 
was planned to definitely measure 
the neutrino oscillation, which was originally suggested by 
the Kamiokande experiment\cite{Kam}.  
An almost pure $\nu_\mu$ beam is generated by the decay of secondary pions
from the KEK 12-GeV proton synchrotron. 
The K2K experiment uses a near detector newly built 
at the KEK site 300~m from the pion production target 
in order to know the beam properties just after its production, 
in addition to the Super--Kamiokande detector\cite{SK},
the far detector 250~km away.  
A muon range detector~(MRD) has been constructed 
as a near detector to measure the energy and angle of muons 
which are produced by charged-current $\nu_\mu N$ interaction. 
This determines the spectrum of the incident neutrino beam.  
The detector also measures the flux and profile of the neutrino beam.  

\section{Design of the muon range detector}

The requirements for the near detector are: 
\begin{enumerate}
\item that it be massive enough to measure the beam flux,
\item that it have large enough transverse dimension to measure
      the beam profile,
\item and that it be capable of measuring the energy of muons 
     in the range of the expected neutrino spectrum, i.e. up to 
     about 3~GeV, with reasonable resolutions.  
\end{enumerate}
To fulfill these requirements, we adopted the range method.  
The MRD consists of 12 layers of iron absorber 
sandwiched between vertical and horizontal 
drift-tube layers.  
The size of a layer is approximately 7.6~m $\times$ 7.6~m.  
In order to have a good energy resolution for the whole energy region, 
the upstream four iron plates are 10~cm thick and the downstream eight are 
20~cm thick.  
The total iron thickness of 2.00~m covers up to 2.8-GeV muons.  

\begin{figure}[htbp]   
\begin{center}
\psfig{file=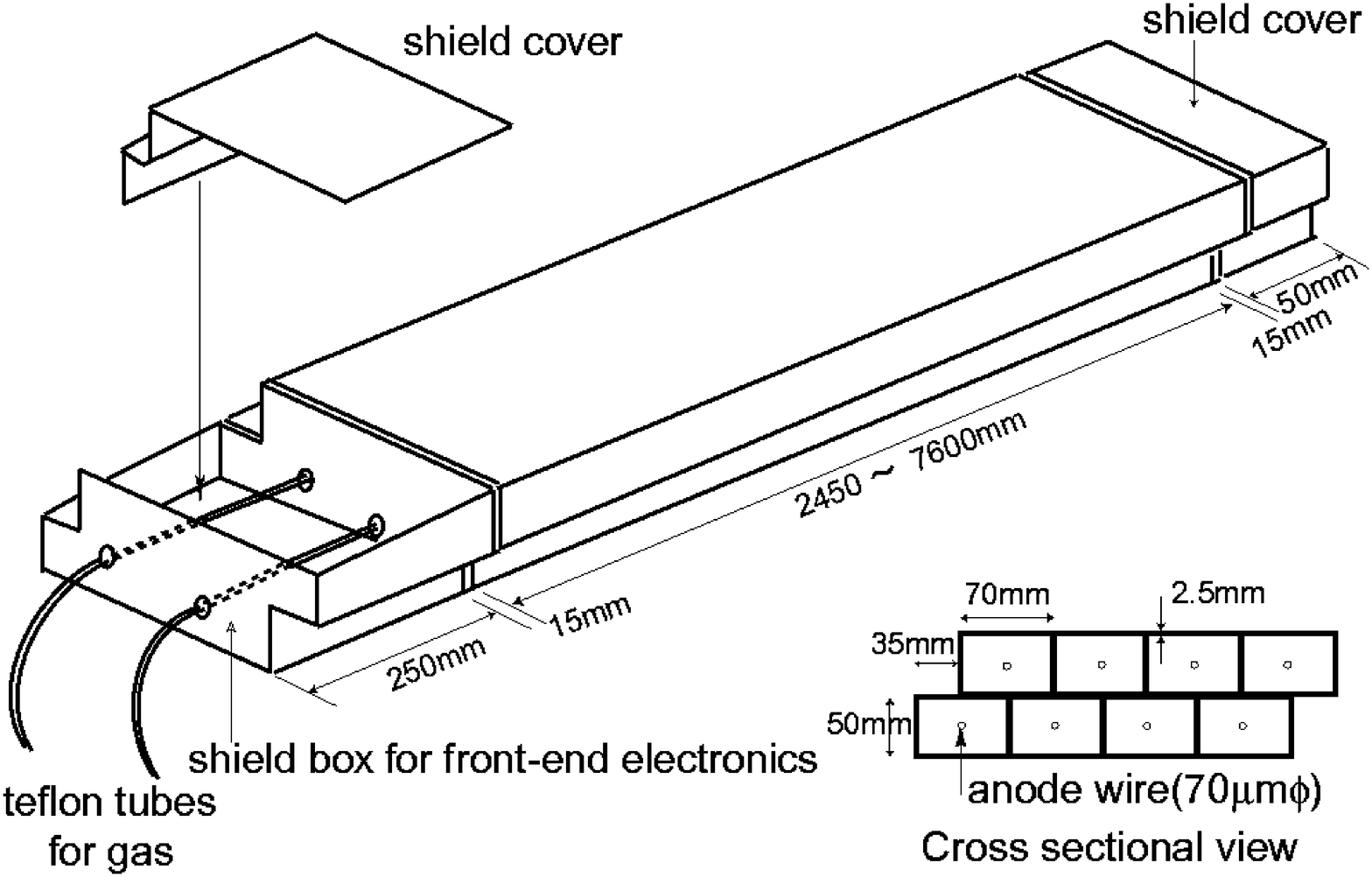,width=11cm}
 \caption{Structure of a drift-tube module.}
 \label{module}
\end{center}
\end{figure}

\begin{figure}[htbp]   
\begin{center}
\psfig{file=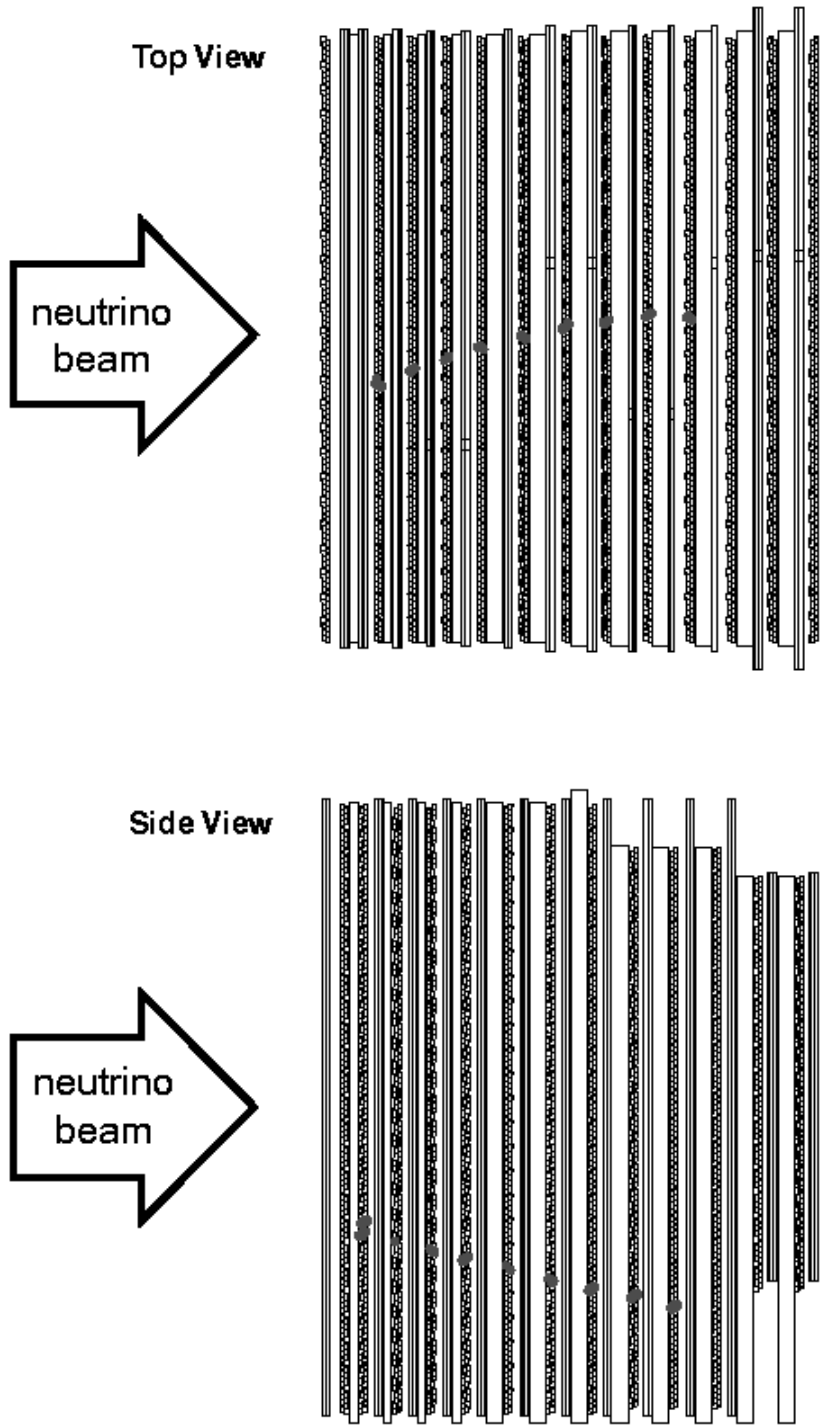,width=11cm}
 \caption{Setup of the MRD.  The top figure shows 
  a top view and the bottom one shows a side view. 
  The overall width and height are each 7.6~m. 
  The iron plates (empty rectangles) are 10 and 20~cm thick. 
  Hits from a typical neutrino-iron event are also plotted.}
 \label{setup}
\end{center}
\end{figure}

The same drift tubes were previously used by 
 the VENUS experiment\cite{VENUSmuc} at TRISTAN.  
Each drift-tube module consists of 8 drift tubes, 
which are arranged as shown in Fig.~\ref{module}.  
The modules vary in length 
from 2.45~m to 7.6~m.  Ideally, we make a drift-tube layer 
by arranging 25 modules of 7.6~m length with a 2-cm gap between 
 consecutive modules.  
However, for some horizontal drift-tube layers 
shorter modules are combined end-to-end 
to make the full 7.6~m $\times$ 7.6~m dimension,
since there are not enough 7.6~m modules. 
We used 829 modules in total:
377 modules of 7.6~m length, and 452 shorter ones.
The available pool from VENUS was 390 and 510 modules, respectively.  
The setup of the MRD 
is shown in Fig.~\ref{setup}.  

The seventh through twelfth iron plates are also from the VENUS 
experiment; the upstream six are new.
The total weight of iron is 864 tons.  
Including the aluminum drift tubes, the total mass 
is 915 tons.  

\section{Tests of the drift tubes}

\subsection{Gas tightness}

\begin{figure}[htbp]   
\begin{center}
\psfig{file=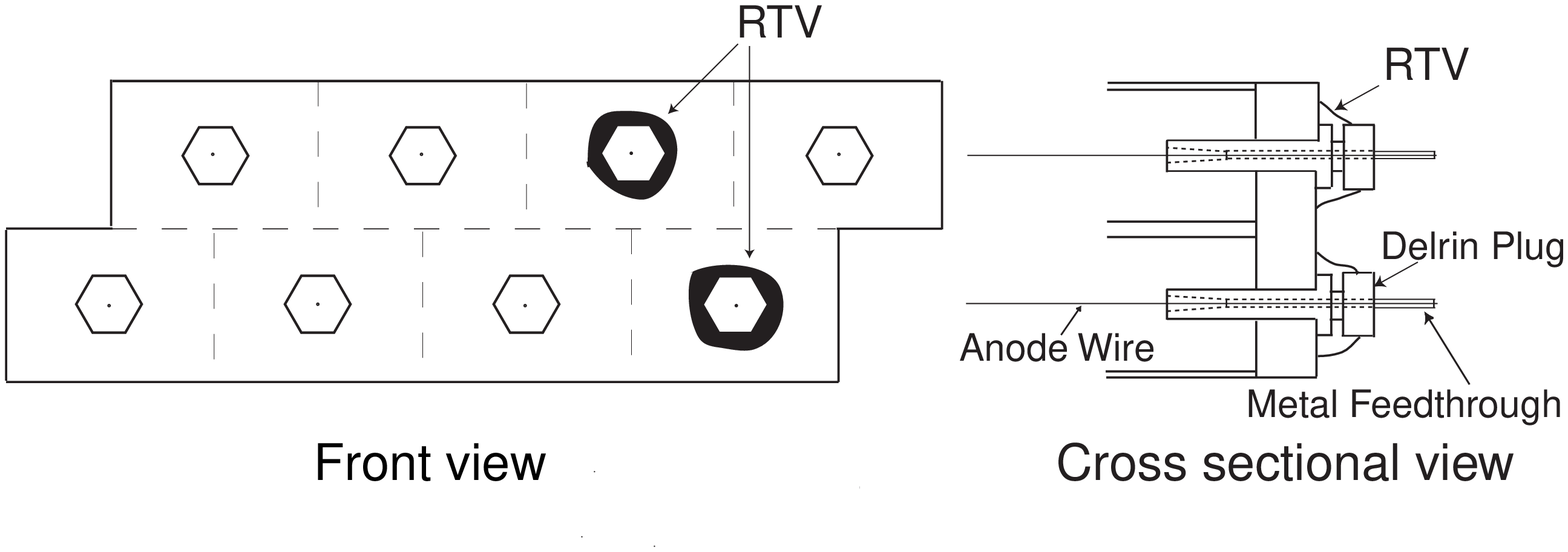,width=11cm}
 \caption{Repair of a gas leak. }
 \label{RTV}
\end{center}
\end{figure}

The gas tightness of the drift tubes was tested by pressurizing them 
to approximately $\rm 1.25~kg/cm^2$ and measuring the decrease 
in the pressure as a function of time.  Nitrogen gas was used for 
this test.  
About 3\% of the modules showed an excessive leak rate corresponding to
more than 1~cc/min under the nominal running conditions.  
Most of the leaks were found at the connection part of the Delrin plug 
to the aluminum end-plate (Fig.~\ref{RTV}).  
Two materials had been used for gas seals by the VENUS group:  
RTV and epoxy glue.  The epoxy glue tends 
to form cracks after a long period.  
Leaking modules were repaired using RTV.
After repair, they showed an average leak rate of less than 0.1~cc/min.

\subsection{Wire tension}

\begin{figure}[htbp]   
\begin{center}
\psfig{file=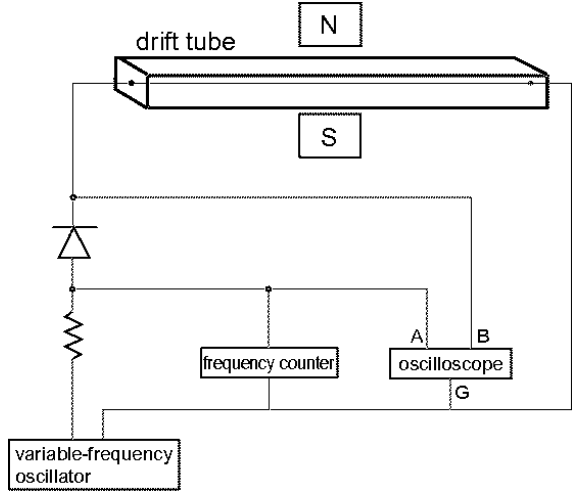,width=7.5cm}
 \caption{Principle of the tension measurement.}
 \label{tensiontest}
\end{center}
\end{figure}

\begin{figure}[htbp]   
\begin{center}
\psfig{file=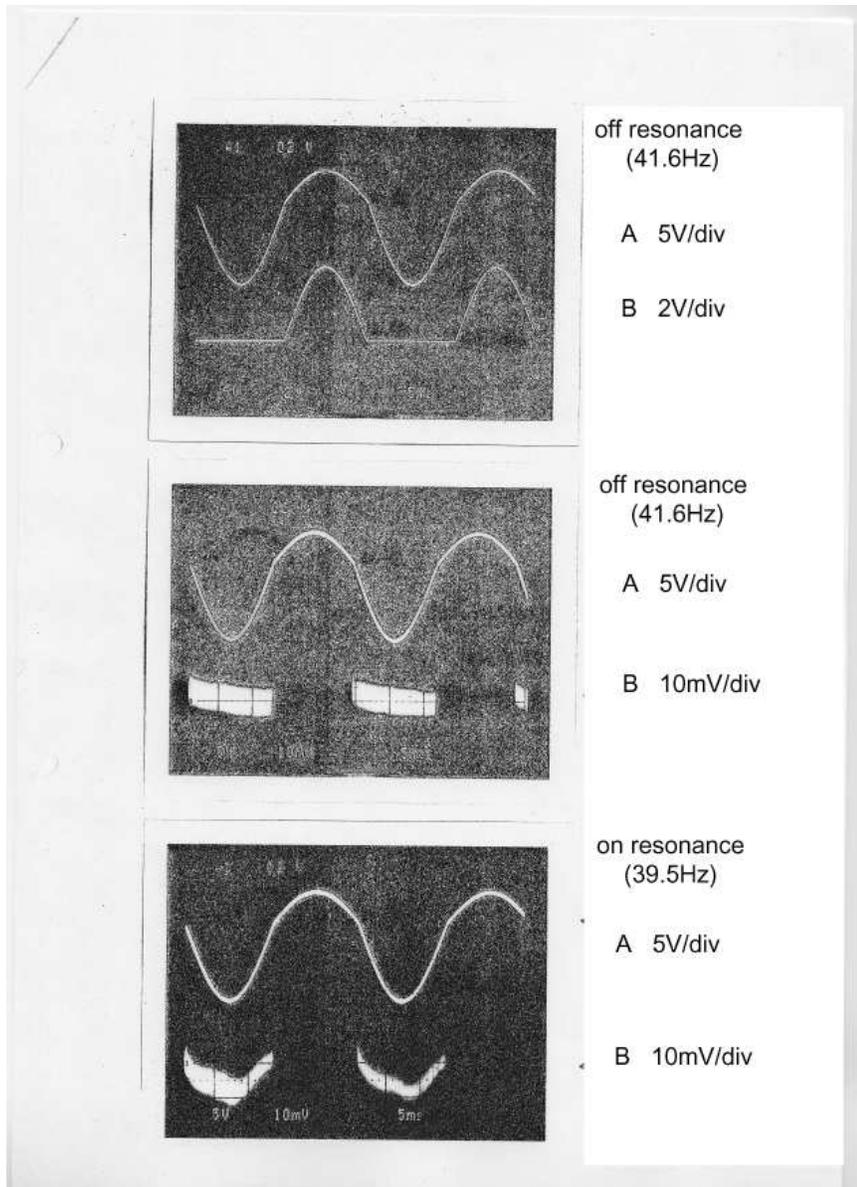,width=11.5cm}
 \caption{Oscilloscope views of the tension test signals. 
  In each picture, channels A and B show the signal at the points 
  A and B of Fig.~\ref{tensiontest}, respectively.  
  The top figure shows signals at a non-resonant frequency.  
  The middle one shows the same, but the scale of the channel B is 
  magnified. 
  The bottom one shows signals at the wire's resonant frequency, 
  where the signal due to the resonant oscillation of the channel B 
  is seen by the curvature of the signal.}
 \label{resonance}
\end{center}
\end{figure}

The wire tension was measured by searching for the resonant frequency 
while supplying a pulsed current in a magnetic field 
(Fig.~\ref{tensiontest}).  
The resonant frequency ($f$ (Hz)) is related to the wire tension ($Mg$ (N)) 
by the following formula: 
\begin{equation}
%\vspace {2ex}
%$f = \frac{n}{2L} \sqrt{\frac{Mg}{m}}$, 
f = \frac{n}{2L} \sqrt{\frac{Mg}{m}} , 
%\vspace {2ex}
\end{equation}
where $n$ is an odd integer, $L$ is the wire length in m, $m$ is
the wire line density (kg/m), and $g = 9.8 \rm m/s^{2}$. 
For this case, the line density of a 70~$\mu$m diameter tungsten wire 
                          is $\rm 7.35\times 10^{-5}$~kg/m.  
The pulsed current was made from a sine wave generated by 
a variable-frequency oscillator, 
which was trimmed by a diode to see a small signal. 
As shown in Fig.~\ref{resonance}, at the resonant frequency, the wire 
started to oscillate in the magnetic field inducing a current, 
which was observed as an order of 10~mV signal at the point B 
of Fig.~\ref{tensiontest}. 

\begin{figure}[htbp]   
\begin{center}
\psfig{file=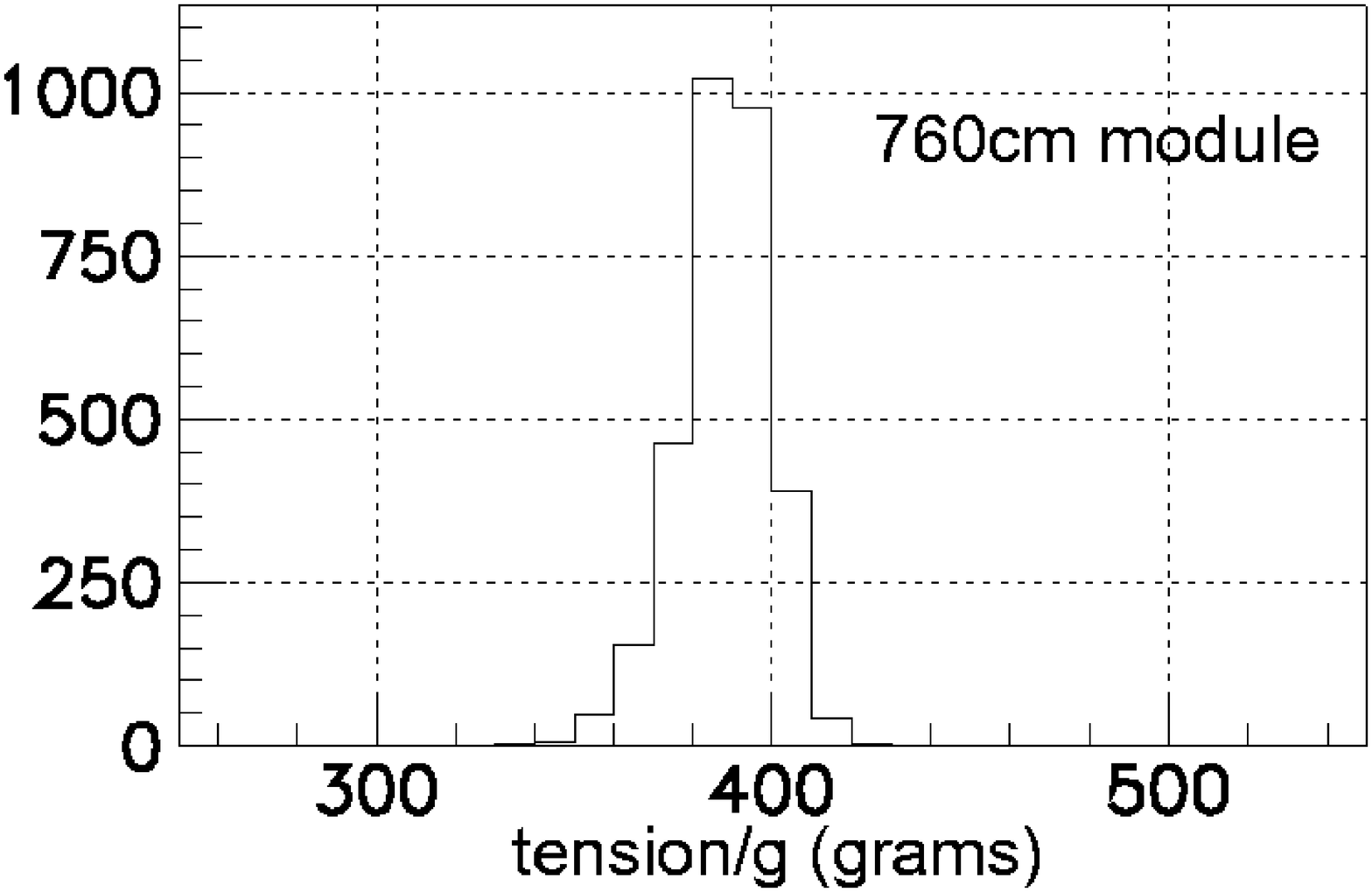,width=8.5cm}
 \caption{Measured tension distribution for the 760-cm-long modules. 
  The data were corrected for $t=20^{\circ}$C.}
 \label{tensiondist}
\end{center}
\end{figure}

The measurement was made by a computer-controlled CAMAC system 
for each of the modules.  
At the same time, the wire resistance was measured as well, 
which is another important check of the continuity of the wires.  
We found 10 broken wires in total out of 7200 wires.  
Most of the broken wires had been cut inside the epoxy glue 
which had been used 
for only some of the modules to fix the wire to the pin.  
The measured tension distribution for the 760-cm-long modules is shown 
in Fig~\ref{tensiondist}.  
The modules with other lengths have similar distributions. 
They are distributed around 390~g with an RMS of about 10~g, which is 
consistent with the VENUS measurement\cite{VENUSten}. 
The broken wires were repaired by stretching a new wire, 
fixed to the pin only by soldering.  

\section{Readout electronics}

\subsection{Modification of the electronics}

\begin{figure}[htbp]   
\begin{center}
\psfig{file=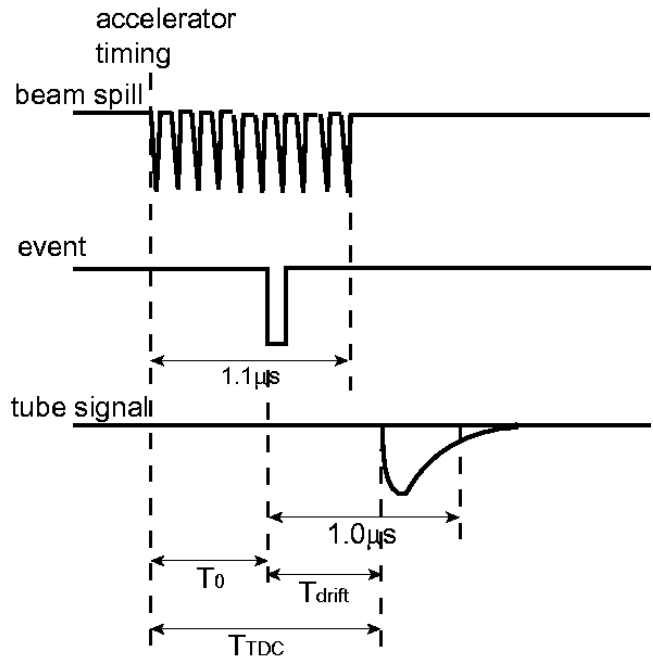,width=8.5cm}
 \caption{Timing chart of the K2K experiment.}
 \label{timing}
\end{center}
\end{figure}

The VENUS electronics for the muon drift tubes\cite{VENUSele} 
consists of single-hit TDCs operated 
in the common start mode.  
The start signal was made from the accelerator timing of every 5~$\mu$sec 
in the TRISTAN experiment. 
The use of a 40MHz 6-bit TDC in the front-end 
electronics boards gives a range 
for the TDC of 1.6~$\mu$sec.  
However, in the K2K experiment, 
the experimental condition is very much different 
from that in the TRISTAN experiment.  A beam spill every 2.2~sec 
lasts for about 1.1~$\mu$sec.  
If we use the accelerator timing as the start signal, 
the TDC range would not cover the sum of the beam spill time and 
the maximum drift time, which is about 2.1~$\mu$sec (Fig.~\ref{timing}). 
Further, the signal from trigger counters is too late to make 
the start signal.  

There were two choices.  One was to modify the electronics to 
operate in a 
common stop mode, so that the trigger counter signal could be used 
for the stop signal.  The other was to change 
the TDC range to be longer than the beam spill time plus the 
maximum drift time, so that we could calculate 
the drift time by subtracting the trigger timing delay later.  
We chose to make the TDC range longer by replacing the 40MHz clock chips 
with 20MHz ones.  
This resulted in a TDC range of 3.2~$\mu$sec.  
This was a simpler modification of the electronics which 
adversely affects only 
the timing resolution slightly, which is not critical for 
the range measurement.  

\subsection{Test of the front-end electronics}

We tested the front-end electronics boards both before and after 
modification.  
Test pulses were injected to the input of the preamplifier on the board. 
The test items were linearity of the stop 
timing and linearity of the start timing, as well as the efficiency 
versus the test-pulse amplitude.
We found about 3\% of the boards to be bad at each stage; 
most of them were repaired.  

\section{Cosmic-ray test of the system}

\begin{figure}[htbp]   
\begin{center}
\psfig{file=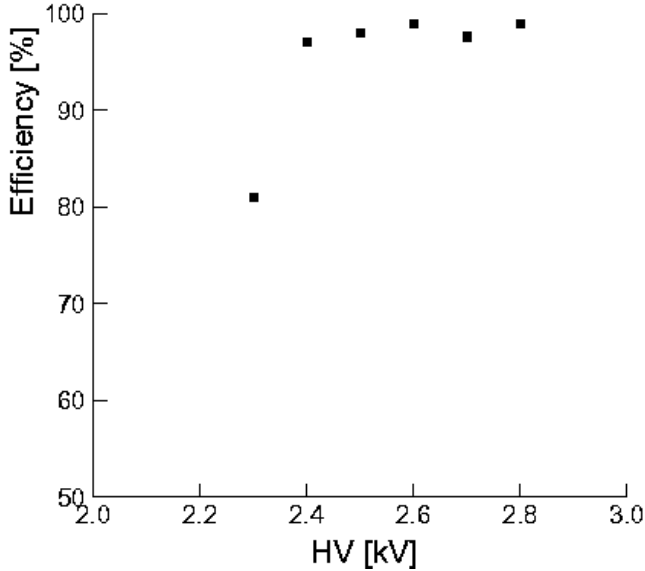,width=8.5cm}
 \caption{Efficiency of the drift tubes as a function of HV.}
 \label{HV}
\end{center}
\end{figure}

A cosmic-ray run was performed to test the complete drift-tube system.  
P10 gas ($\rm Ar:CH_{4}=90\%:10\pm0.5\%$) was flowed 
as in the real experiment.  
The results of the cosmic-ray test show that we have an efficiency 
plateau    above 
2.5~kV of high voltage (Fig.~\ref{HV}), and 
a spatial resolution of about 2~mm.
We chose the nominal operating high voltage to be 2.7~kV.  

The time-to-distance relation was also investigated. 
This study confirmed the VENUS parameterization of the relation:
\begin{equation}
%\vspace {2ex}
x = 2.11\times t^2 + 2.72\times t , 
%$x = 2.11\times t^2 + 2.72\times t$, 
%\vspace {2ex}
\end{equation}
where $x$ is the drift length in cm and $t$ is the drift time 
in $\mu$sec.  

\section{Gas system}

\begin{figure}[htbp]   
\begin{center}
\psfig{file=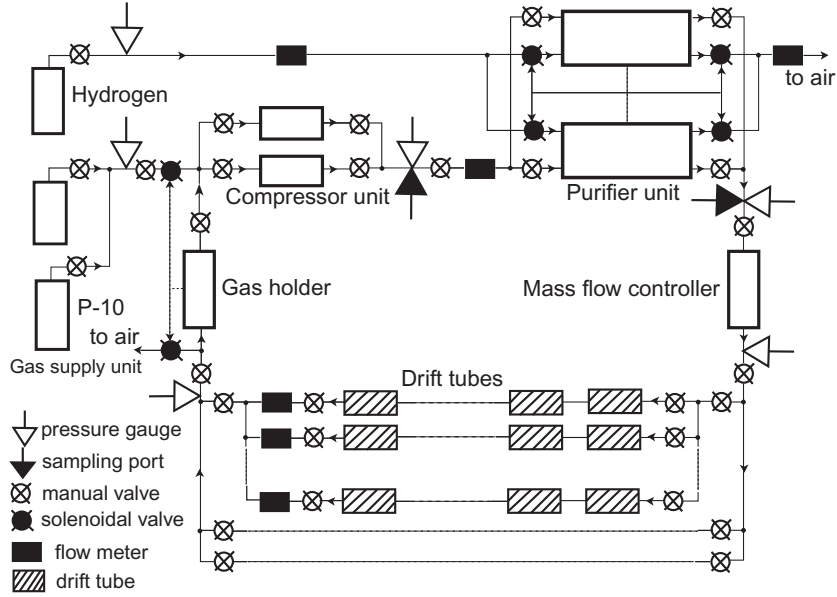,width=11cm}
 \caption{Block diagram of the gas recirculation system.}
 \label{gas_system}
\end{center}
\end{figure}

In the K2K running, P10 gas is supplied to the drift tubes using 
the gas recirculation system of the VENUS experiment.  
As is shown in Fig.~\ref{gas_system}, 
it is composed of a gas-supply unit, a compressor, a purifier 
and a gas holder. 
The compressor and the purifier each has a duplicate for backup. 
P10 gas is pressurized by the compressor and then sent to the purifier.  
The purifier includes a reaction column and an adsorption column. 
In the reaction column, oxygen contamination is trapped by flowing 
the gas through porous nickel metal.  In the adsorption column, 
water contamination is removed using molecular sieves.  
The purified gas is flowed to the drift tubes at a rate of $50~l$/min. 
At this flow rate, the total volume of the drift tubes of 130~$\rm m^{3}$ is 
replaced every 2 days.  
Returned gas from the drift tubes is stored in a gas holder, 
which keeps the gas pressure 0.01~$\rm kg/cm^{2}$ higher than 
 atmospheric pressure.
The gas in the gas holder is sent to the compressor and recirculated. 

Two sampling ports are located at the input and the output
of the purifier. 
Sampled gas is measured by gas chromatography on a weekly basis 
to check the purity.  Normally, the oxygen contamination of 
the return gas is less than 100ppm, and that of the purified gas is 
less than 5ppm.  
When the contamination level becomes higher, we regenerate the purifier.  
To this extent, P10 gas mixed with hydrogen is flowed through 
the reaction column heated to $150 - 200^{\circ}$C 
in order to deoxidize the nickel catalyst.  The molecular sieves 
are also heated to $250 - 300^{\circ}$C for regeneration.  

\section{Iron absorber}

\begin{figure}[htbp]   
\begin{center}
\psfig{file=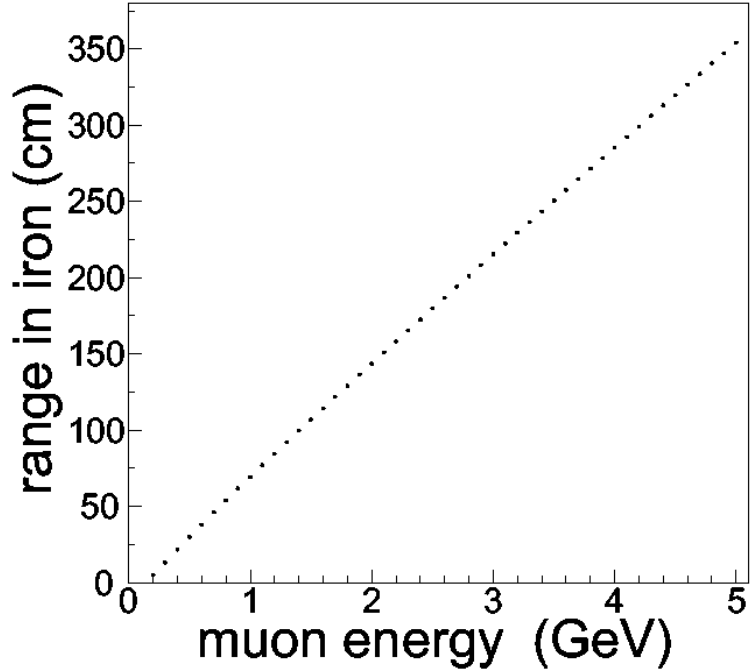,width=10cm}
 \caption{Muon range in iron vs. muon energy calculated using the GEANT code.}
 \label{range}
\end{center}
\end{figure}

Iron plates are used not only as an absorber for range measurements, 
but also as a target for the neutrino beam.  
Accurate knowledge of the thickness of the plates is necessary 
for these measurements.  
It is guaranteed by the Japanese Industrial Standard (JIS) 
that the thickness of the upstream 4 plates is $100 \pm 1.7$~mm 
and that of downstream 8 plates is $200 \pm 2$~mm.  
The relationship between the muon energy and the muon range 
in     iron was calculated using a GEANT based Monte Carlo code\cite{GEANT}.  
The result is shown in Fig.~\ref{range}.  

\section{Acceptance and expected resolutions}

The acceptance and resolutions of the MRD were studied 
by a Monte-Carlo method. 
In the simulation, 
muons with various energy were injected in the forward direction 
from randomly chosen vertices in a radius of 
1m of the 1st iron plate. 
They were simulated in the detector 
considering both ionization and multiple scattering.  
The hit positions were converted into drift times.  
The muon track was reconstructed using the tracking algorithm 
explained in the next section.  

\begin{figure}[htbp]   
\begin{center}
\psfig{file=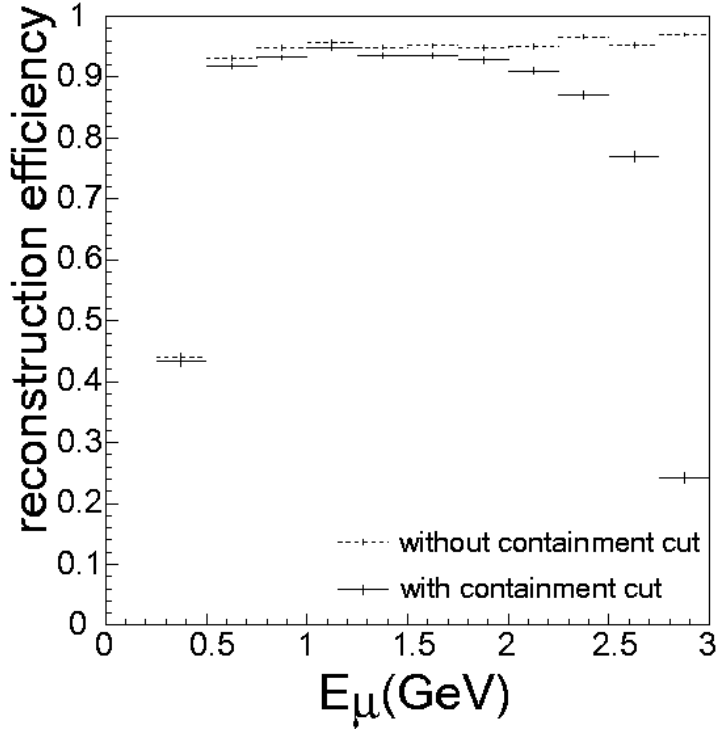,width=9.5cm}
 \caption{Reconstruction efficiency as a function of the muon energy. 
  The dotted bars show the efficiency without the containment cut. 
  The solid-line bars show the efficiency with the containment cut. 
  The vertical bars show the statistical error of the Monte-Carlo 
  calculation and the horizontal bars show the bin size.}
 \label{receff}
\end{center}
\end{figure}

The reconstruction efficiency is plotted as a function of 
the muon energy in Fig.~\ref{receff}.  
In some analyses, we require a track to be contained in the detector 
to make an energy measurement possible.  
Tracks which exit the detector are not contained. 
The efficiency was derived both with and without the containment cut.  

\begin{figure}[htbp]   
\begin{center}
\psfig{file=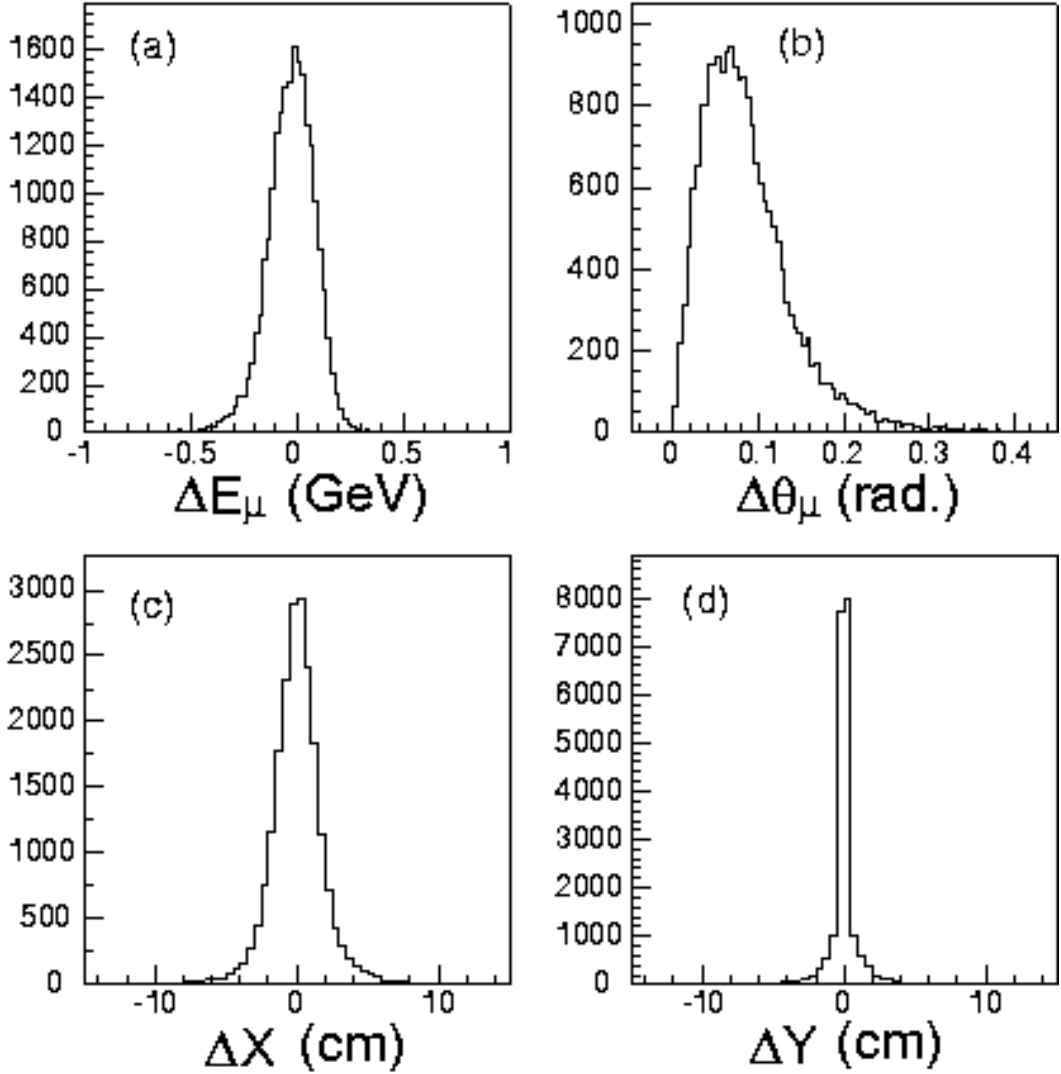,width=14cm}
 \caption{Expected resolutions of the MRD. 
  a) energy resolution; b) angular resolution; 
  c) horizontal vertex resolution and d) vertical vertex resolution. 
  The vertical vertex resolution is better than the horizontal one 
  because the horizontal tubes are closer to the 1st iron layer.}
 \label{reso}
\end{center}
\end{figure}

The differences in the reconstructed and generated muon energies 
are plotted in Fig.~\ref{reso}-a, which is taken as an energy resolution.  
The same is plotted for the muon angle, as well as the horizontal and 
vertical start positions in Figs.~\ref{reso}-b, c and d, 
respectively.  

\section{Tracking algorithm}

As stated above, one drift-tube module consists of two 
half-cell staggered planes.  The tracker treats each layer as 
two independent drift-tube planes.  
The tracking proceeds in the following 4 steps.  
In steps 1, 2 and 3 the x-z view and the y-z view are treated 
independently, where the z-axis is defined by the beam direction 
and the y-axis is defined by the upward direction.  
\begin{enumerate}
\item Cell fitting: in this step, the drift time is ignored and 
the wire positions are used as hit positions.  
A straight line is drawn between every combination of two hits 
on different planes which are apart by more than three planes.  
The number of hits near the line (typically closer than 7~cm) is counted.  
If the number of hits is large enough 
(typically larger than 60\% of the number of planes 
traversed by the track), it is regarded as a cell-track.  
\item Fragment fitting: based on a cell-track, up to six successive 
planes are examined.  If there exist 3 hits, it is regarded as 
a fragment, and a linear fit is done taking into account of 
the drift times and left/right ambiguity.  
In this fitting $T_{0}$, the interaction time,
                          is also varied as one of the parameters.  
Every other plane in a cell-track is taken 
as the start point of a fragment.  
\item 2D fitting: the most upstream fragment is taken as the start 
of a 2D-track.  A fragment which can be smoothly connected to 
the previous fragment is searched for.  
Typical connection criteria are that the distance between the two 
lines is shorter than 10~cm and difference of the two slopes 
is less than 0.15.  This process is iterated until no
connecting fragment can be found.
This group of connected fragments is 
called a 2D-track.  Used fragments are removed from the list, 
and the process is repeated for the remaining fragments.  
\item 3D fitting: the overlap in the z-direction is examined between 
the xz 2D-tracks and the yz 2D-tracks. 
The pair of (xz,yz) 2D-tracks which has the longest 
overlap is taken as a 3D-track.  The used 2D-tracks are removed from the list 
and the process is repeated for all the remaining 2D-tracks.  
\end{enumerate}

As a result, the shortest track reconstructed consists of one
fragment which has 
three successive hits in both the xz and yz views.  
This track traverses at least one iron plate corresponding to 
an energy threshold of 150~MeV (300~MeV) for a 10-cm (20-cm) thick iron plate. 

\section{Performance of the muon range detector}

The MRD measures muons produced by the charged-current neutrino 
interaction in the iron plates.  
A typical neutrino-iron interaction event is shown in Fig.~\ref{setup}.  

\subsection{Tracking efficiency and the hit efficiency}

\begin{figure}[htbp]   
\begin{center}
\psfig{file=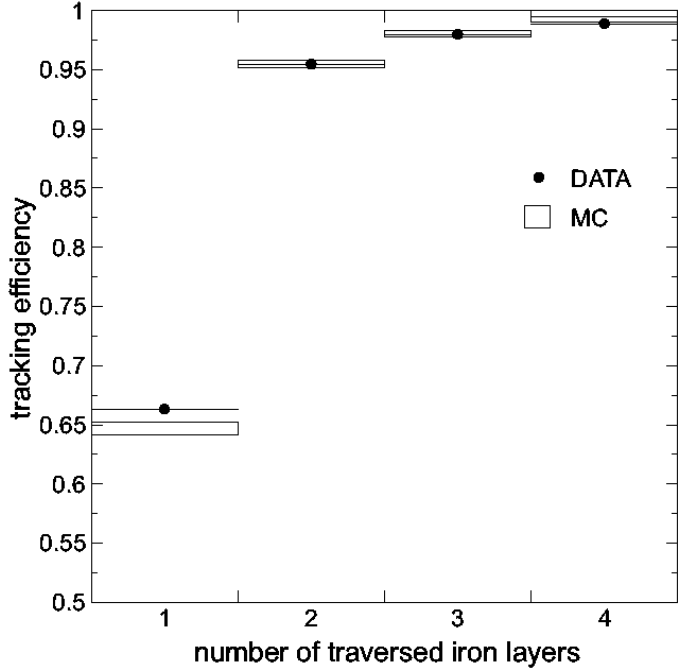,width=9cm}
 \caption{Tracking efficiency as a function of the traversed iron layers. 
  The boxes show the Monte-Carlo result, with a size representing the 
  errors.}
 \label{trackeff}
\end{center}
\end{figure}

The tracking efficiency as a function of the track length 
(number of traversed layers) 
is derived as follows using beam data: 
\begin{enumerate}
\item Tracks longer than the test region by at least one iron layer 
      both upstream and downstream 
      are selected as sample tracks  in order to assure the existence 
      of a track in the test region.  
\item All hits except for those in the test region are masked.  
\item The masked data are fitted by the tracker to see whether 
      the tracker finds the track or not.  
\end{enumerate}
The tracking efficiency is dependent on the hit efficiency of 
the drift tubes.  We determined the effective hit efficiency 
so that the Monte-Carlo result on the tracking efficiency would reproduce 
the beam data well.  The obtained effective hit efficiency is 97.5\%. 
The inefficiency of 2.5\% includes the geometrical effect 
coming from a possible mis-alignment of the drift tube 
modules, as well as the real hit inefficiency.  
The result of the tracking efficiency is shown in Fig.~\ref{trackeff} 
with the Monte-Carlo result overlaid.  

\subsection{Noise rate}

The noise rate was estimated from off-spill data, which are taken 
in the time between successive beam spills.  These are one class of 
randomly triggered data.  The average noise hits 
in the off-spill data amounts to approximately 27 out of 6632 total tubes, 
implying a 0.4\% average noise rate.  This rate is composed of all the 
non-beam-related background hits, including cosmic-rays.  
The main part of this comes from electronic noise, most of which is 
easily cut using timing information.  
As a result, noise hits do not change the analyzed results.  

\subsection{Detector stability}

\begin{figure}[htbp]   
\begin{center}
 \psfig{file=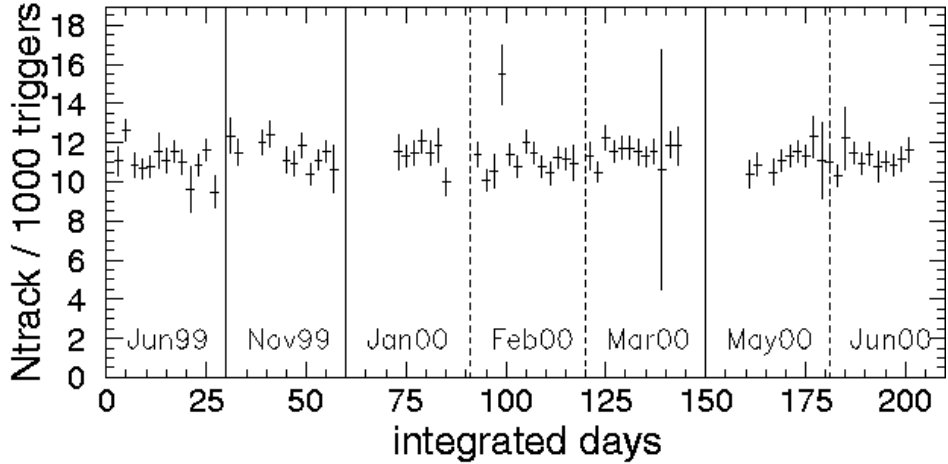,width=12.5cm}
 \caption{Average number of reconstructed 3D-tracks from the off-spill 
  data as a function of time.}
 \label{offrate}
\end{center}
\end{figure}

\begin{figure}[htbp]   
\begin{center}
 \psfig{file=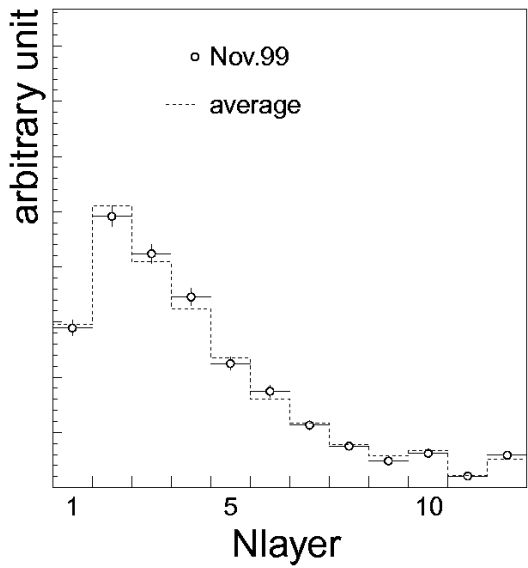,width=7cm}
 \caption{Distribution of number of traversed layers by a cosmic-ray track  
  in November 1999 (data points) compared with the averaged distribution
  over all experimental periods (dashed histogram). 
  Corresponding plots for the other one-month periods show similar 
  agreement.}
 \label{offntrack}
\end{center}
\end{figure}

\begin{figure}[htbp]   
\begin{center}
 \psfig{file=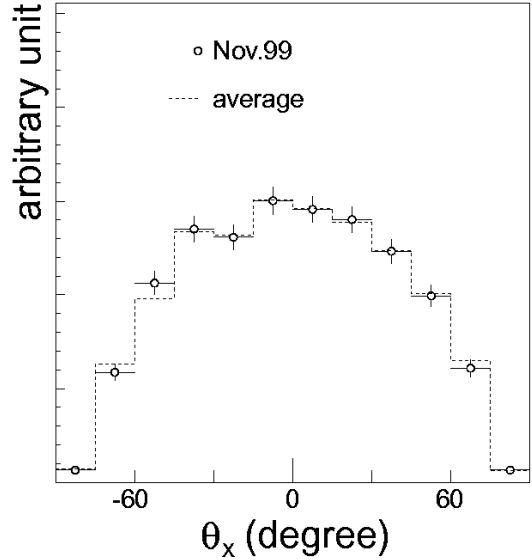,width=7cm}
 \caption{Angular distribution of cosmic-ray tracks in the x-z view 
  in November 1999 (data points) compared with the averaged distribution
  over all experimental periods (dashed histogram).}
 \label{offthetax}
\end{center}
\end{figure}

\begin{figure}[htbp]   
\begin{center}
 \psfig{file=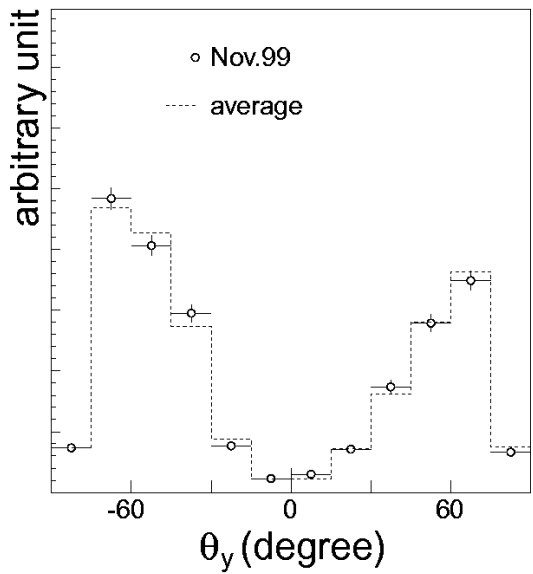,width=7cm}
 \caption{Angular distribution of cosmic-ray tracks in the y-z view 
  in November 1999 (data points) compared with the averaged distribution
  over all experimental periods (dashed histogram). 
  They have a two-peak structure because the tracker 
  assumes all tracks start from upstream 
  with respect to the beam (lower z value).}
 \label{offthetay}
\end{center}
\end{figure}

As of June 2000, 
the MRD has been alive 91.7\% of the time since the start of 
the K2K data taking in June 1999. 
The off-spill data were analyzed to study the detector stability. 
Only cosmic-ray muons contribute to the off-spill data as tracks. 
The average number of reconstructed 3D-tracks is plotted 
as a function of time in Fig.~\ref{offrate}. 
The average number of tracks is about 11/1000 triggers and 
has been stable within the statistical error. 
The number of traversed layers by a track is correlated with the energy 
loss of the particle in the MRD.  This distribution is analyzed 
on a monthly basis.  
The distribution in November 1999 is compared with 
the averaged distribution over all experimental periods 
in Fig.~\ref{offntrack}. 
Distributions in other months are the same within statistical errors. 
Angular distributions, also made on a monthly basis, in the x-z and 
y-z views are shown for the November 1999 and total data sets 
in Figs.~\ref{offthetax} and ~\ref{offthetay}, 
respectively.  
Distributions in other months are same within statistical errors 
for both angular distributions.  
The reproducibility of these distributions shows 
that the detector has been working 
stably in all experimental periods. 

\section{Summary and conclusion}

The MRD was constructed as a near detector of 
the K2K long-baseline neutrino experiment.  
It measures the position, momentum, and direction of
 muons produced in charged-current neutrino interactions.  
The coverage of the detector is $\pm12$~mrad with respect to 
the production target, and 
the fiducial mass is 329 tons.  
The energy acceptance is from 0.3~GeV to 2.8~GeV with a resolution of 
0.12~GeV for forward-going muons.  
The track angular resolution is about 0.09~rad and the vertex resolution 
perpendicular to the beam direction about 2~cm.  

The MRD has been working stably since the start of the K2K experiment.  

\begin{ack}

We gratefully acknowledge the cooperation of all other K2K members 
for this work. 
We appreciate the VENUS group for allowing us to reuse 
their muon system. 
Especially we wish to thank Dr. Y. Asano for sharing with us his 
experience with the system.  

This work has been supported by the Japanese Ministry of 
Education, Science, Sports and Culture (the Monbusho), 
its grants for Scientific Research, 
the U.S. Department of Energy, 
the Korea Research Foundation Grants and 
the Korea Science and Engineering Foundation. 
\end{ack}


\begin{thebibliography}{9}
\bibitem{K2K} K2K Collaboration, to be published in Phys. Lett. {\bf B} (2001); 
K. Nishikawa {\it et al.}, KEK-PS proposal (E362) (1995).
\bibitem{Kam} K. S. Hirata {\it et al.}, Phys. Lett. {\bf B 205} (1988) 416.
\bibitem{SK}  Y. Fukuda {\it et al.}, Phys. Rev. Lett. {\bf 81} (1998) 1562.
\bibitem{VENUSmuc} Y. Asano {\it et al.}, Nucl. Instr. and Meth. 
{\bf A 259} (1987) 430.
\bibitem{VENUSten} Y. Asano {\it et al.}, Nucl. Instr. and Meth. 
{\bf A 254} (1987) 35.
\bibitem{VENUSele} Y. Ikegami {\it et al.}, IEEE Trans. on N. S. 
{\bf 36} (1989) 665.
\bibitem{GEANT} R. Brun {\it et al.}, CERN DD/EE/84-1 (1987).
\end{thebibliography}
\end{document}